\documentclass[final]{siamltex}
\usepackage{graphicx}
\usepackage{bm}
\usepackage{subfigure}

\title{Percolation-induced exponential scaling in the large current tails of random resistor networks}
\author{Feng Shi \thanks{Department of Mathematics,University of North Carolina at Chapel Hill, Chapel Hill, NC 27599-3250} \and
        Simi Wang \footnotemark[1] \and
        M. Gregory Forest \footnotemark[1] \and
        Peter J. Mucha \footnotemark[1]}

\begin{document}
\maketitle
\begin{abstract}
There is a renewed surge in percolation-induced transport  properties of diverse nano-particle composites (cf.  RSC Nanoscience \& Nanotechnology Series, Paul O'Brien Editor-in-Chief). We note in particular a broad interest in nano-composites exhibiting sharp electrical property gains at and above percolation threshold, which motivated us to revisit the classical setting of percolation in random resistor networks but from a multiscale perspective. For each realization of random resistor networks above threshold, we use network  graph representations and associated  algorithms to identify and restrict to the percolating component, thereby preconditioning the network both in size and accuracy by filtering  {\it a priori} zero current-carrying bonds. We then simulate many realizations per bond density and analyze scaling behavior of the complete current distribution supported on the percolating component. We first confirm the celebrated power-law distribution of small currents at the  percolation threshold, and second we confirm results on scaling of the maximum current in the network that is associated with the backbone of the percolating cluster. These properties are then placed in context with global features of the current distribution, and in particular the dominant role of  the large current tail that is most relevant for material science applications. We identify a robust, exponential large current tail that: 1. persists above threshold; 2.  expands broadly over and dominates the current distribution at the expense of the vanishing power law scaling  in the small current tail; and 3. by taking second moments, reproduces the experimentally observed power law scaling of bulk conductivity above threshold.
\end{abstract}



\section{Introduction}
Motivated by open questions associated with nano-composite materials engineering \cite{bauhofer_review_2009, bergman_physical_1992, byrne_recent_2010, nan_physical_2010, white_electrical_2010}, we revisit the classical problem of bond percolation \cite{kirkpatrick_percolation_1973} in a three-dimensional $L\times L\times L$ cubic lattice. In particular, we are interested in the relationship between macroscopic conductivity scaling behavior and the scaling of the underlying current distributions on the percolating bonds. In current multiscale terminology, we seek to relate the statistical scaling properties of the bond-scale currents with the macroscopic measurable scalar quantity, the bulk conductivity. Percolation theory  in the idealized model of cubic lattices \cite{deutscher_percolation_1983, grimmett_percolation_1999, kesten_percolation_1982, sahimi_applications_1994, stauffer_introduction_1994} has been very successful in explaining the most salient experimental phenomena, in particular the power-law scaling of conductivity near percolation threshold \cite{bauhofer_review_2009, bergman_physical_1992,  byrne_recent_2010, clerc_electrical_1990}.

For most composite materials, where a low-performing host matrix is endowed with a high-performing particle phase, the geometry of particle-phase contacts is not a cubic lattice, and the properties at the bond or contact sites are often non-uniform.  Yet for realistic composite materials, the classical lattice-bond percolation results have provided excellent qualitative agreement with near-threshold, macroscopic behavior.  Two natural considerations (detailed below) arise with the engineering emphasis of the past two decades on nano-particle composites \cite{balberg_excluded_1984, bauhofer_review_2009, bergman_physical_1992, berhan_modeling_2007, bug_continuum_1985, byrne_recent_2010, foygel_theoretical_2005, nan_physical_2010, neda_reconsideration_1999, white_electrical_2010}. High aspect ratio particles (rods or platelets) are dispersed in traditional polymers, which percolate at low volume fractions. Even though the particle contacts are randomly distributed in space, and the contacts vary with particle-particle angle and position of contact, the bulk scaling behavior of conductivity at and near percolation threshold is qualitatively captured by percolation theory of cubic lattice models.

The two considerations noted above are the following.  First, the nano-phase is typically expensive and the aim is to use volume fractions above but close to percolation threshold.  This design constraint necessarily calls into question the stability of the material performance to perturbations in the percolating particle phase.  We address this question in the current paper using network graph methods to identify the percolating component of bonds, and then to analyze the statistical properties of the currents passing through the percolating component.  We then show that the bulk measurable conductivity, and its scaling behavior, results from the bond-scale properties and scaling behavior of the statistical current distributions, by taking second-moments of the current distributions.  The second consideration has to do with randomizing the cubic lattice assumption to be consistent with the contacts between high aspect ratio rods or platelets.  This requires analysis of actual 3D rod dispersions instead of cubic lattices, which we defer to another study \cite{shi_network_2013}. The reader is also referred to recent numerical studies of rod-like dispersions where the sensitivity of the threshold volume fraction of rods to finite aspect ratio and polydispersity in aspect ratio are explored \cite{white_electrical_2010}.

Here we aim to contribute to the understanding of current distributions across the bond percolation network which, although constrained on a cubic lattice, can shed lights on current distributions and multiscale properties in continuum systems. In the seminal work of de Arcangelis \emph{et al.} \cite{de_arcangelis_anomalous_1985, de_arcangelis_multiscaling_1986}, a hierarchical lattice model is given for the percolating backbone of the network, which yields a log-normal current distribution in the network; this model was later generalized by Lin \emph{et al.} \cite{lin_multifractal_1991}. These models successfully capture the multifractal behavior of the current distribution \emph{at percolation threshold} (e.g., an infinite hierarchy of exponents in the moments).  However, they fail to predict the power law distribution of small currents (Straley \cite{straley_current_1989} and Duering \emph{et al.} \cite{duering_current_1990, duering_current_1992}), and they do not address additional features of the current distribution that are most relevant to materials applications (specifically, the large current distribution properties as discussed in the following).

Current distributions and their scaling behavior, if they can be quantified, have fundamental importance in materials science. Low moments of the current distribution dictate physically measurable, macroscopic  properties that have been the principal target of homogenization and percolation theory. Specifically, the second moment of the current distribution yields the bulk conductance, while the fourth moment is relevant to the flicker ($1/f$) noise of the system \cite{de_arcangelis_anomalous_1985,rammal_flicker_1985}. Therefore, measurable macroscopic scaling behavior at or above percolation threshold (cf. \cite{nan_physical_2010}) is inherited from the current distribution.  Another illustration is in the study of breakdown of random media \cite{batrouni_fracture_1998, de_arcangelis_random_1985, duxbury_breakdown_1987, kahng_electrical_1988}. This critical network property motivated studies on the size and location of the largest current in the network \cite{chan_large_1989, kahng_logarithmic_1987, li_size_1987, machta_largest_1987}. Li and Duxbury \cite{li_size_1987} showed that the logarithmic scaling of the largest current with respect to the system size is consistent with an exponential tail of the current distribution. Chan {\em et al.} \cite{chan_large_1989} showed that large currents in a ``funnel-shaped" region have an exponential distribution. As we shall show in the following sections, averaged (bulk) network properties and a one-point statistic such as the largest current in the network are not only implied, but directly controlled by, the large current tail of the distribution.

With knowledge on the very small currents and the largest current, little is known or has been reported on the bridge between the two ends of the current distribution. Indeed, our study reveals the power law scaling at percolation threshold of the small current tail of the current distribution is not responsible for the experimentally observed power law scaling in bulk conductivity nearby yet above threshold.  Here we analyze the entire current distribution, reproducing the power-law distributions of small currents \cite{duering_current_1990, duering_current_1992,  straley_current_1989} while revealing the dominating extent of an exponential large current tail. We then take moments to show the robustness of the exponential large current tail above criticality, which is independent of the bond density given a unit uniform electric field in the system. We use the term ``robustness" to emphasize that the exponential tail is insensitive to bond density above (but relatively close to) percolation threshold and that the gradients of the best-fit exponential rates of the large current tail are smooth with respect to bond density variations.  Lastly we show how the exponential current tail controls and dictates the scaling behavior of the largest current and the power-law scaling of the bulk conductivity. We point the reader to Figure \ref{scaling2} in particular for an illustration of robustness with respect to bond density above the critical percolation threshold, not only in the exponential rate of the large current tail, but also in the bulk conductance of the network and in the percentage of significant current-carrying bonds. These connections between multiscale electrical properties of the network motivate the methods developed and applied herein.

\section{Model and Method}
A random resistor network is an $L\times L\times L$ cubic lattice in which each edge of the lattice takes conductance $1$ with probability $p$ (it is traditional to call such a conducting edge a bond and $p$ is the bond density), and conductance $0$ with probability $1-p$. We seek to understand the relationship between the electrical and topological properties of the resulting bond percolation network. In order to model an externally-driven bulk electrical response, we consider two perfectly conducting $L\times L$ plates to be present at opposite ends of the cube, representing the sink and source of current (in response to either an external voltage drop or current source).

The bond percolation threshold $p_c$ for an infinite 3D cubic lattice is $p_c\doteq0.2488$ \cite{lorenz_precise_1998}. For $p<p_c$, all clusters are small and almost surely no percolating cluster forms in an infinite network. At $p_c$ an infinite cluster emerges with finite non-zero probability that  spans the network, i.e., the network has a percolating cluster. There is a significant literature devoted to the scaling behavior of the distribution of cluster sizes. The typical representation is in terms of a power law with a exponential cutoff \cite{stauffer_introduction_1994},
\begin{displaymath}
 n_s\propto s^{-\tau}\exp(-|p-p_c|^{1/\sigma}s),
\end{displaymath}
where $n_s$ is the number of clusters with $s$ bonds. Nonetheless, there is no known connection between the cluster distribution scaling and the distribution of currents supported on the cluster distribution. Indeed, the non-zero values of the current distribution are associated with the geometric properties of only those clusters that percolate, whereas all other clusters are lumped together in the zero current value.  Clearly, some connection exists between this cluster size scaling and the small and large tails of the current distribution, yet this remains an open problem.

With this background, the simulation procedure is now summarized. We solve for the physical distribution of currents by large-scale simulation of the random resistor network. Specifically, for each realized graph of the random resistor network model in which nodes correspond to the lattice points and edges to the conducting bonds, a breadth-first search algorithm \cite{knuth_art_1997} identifies the union of percolating clusters that connect the two plates. The key ingredient is the plate-constrained 2-core --- defined here as the connected subgraph containing both boundary plates with degree at least two in the interior of the subgraph. {\em This plate-constrained 2-core captures all bonds that potentially carry non-zero current in the posed problem}. This pre-processing step provides two significant advantages. 1.\ Restriction to the plate-constrained 2-core filters approximately 90\% of the bonds near percolation threshold, therefore reducing the linear system to $10\%$ of its original size. 2.\ By {\it a priori} elimination of all bonds not in the plate-constrained 2-core, we remove a vast fraction of exactly zero-current bonds from the numerical simulation of the linear system, thereby improving numerical precision overall, and in particular improved resolution of the small current tail. Kirchhoff's law \cite{strang_introduction_1986} is then solved on the plate-constrained 2-cores with a standard linear solver, giving the current on each bond. A statistical description of the network properties is obtained by averaging over $1000$ realizations for each bond density and system size.

\section{Results and Discussions}

\subsection{Multiscale Current Distributions}
Because it is extremely rare that a bond in the percolating backbone carries exactly zero current, we separate bonds according to zero current (within numerical precision) from all the rest, and analyze the non-zero currents. Let $f(i)$ denote the probability density function (PDF) of the currents across the population of current-carrying bonds (that is, ignoring zero-current bonds where present). Let $h(x)$ be the corresponding PDF (again, restricted to non-zero currents) of the logarithmic current $X=\ln(I)$. The two distributions are related by
    $h(x)= f(e^x)e^x.$

\begin{figure*}
  \begin{minipage}{0.5\textwidth}
    \includegraphics[width=\textwidth]{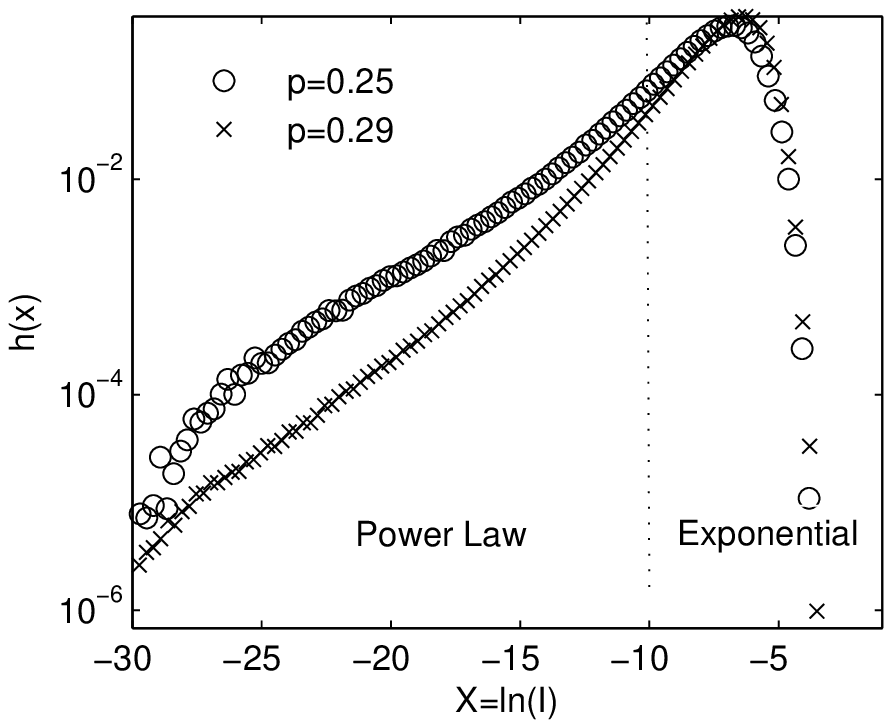}
  \end{minipage}%
  \begin{minipage}{0.5\textwidth}
    \includegraphics[width=\textwidth]{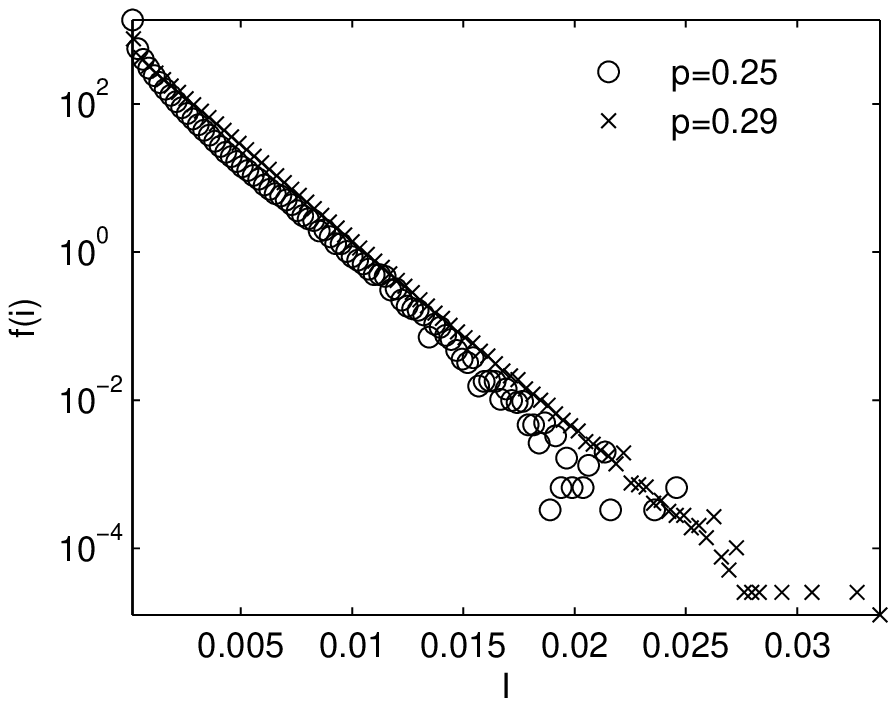}
  \end{minipage}
  \caption{Distribution h(x) of the logarithmic current (left) and distribution f(i) of the current (right) near threshold ($p=0.25$) and above threshold ($p=0.29$) in $100\times100\times100$ random resistor networks. The logarithmic transformation of currents exposes small currents which have a power law distribution (left panel) while the overall current distribution looks exponential (right panel).}\label{hxfi}
\end{figure*}

The logarithmic current distribution $h(x)$ and current distribution $f(i)$ near ($p=0.25$) and above ($p=0.29$) the threshold are shown in Figure \ref{hxfi} {\em for a unit voltage source}. (Below we consider the alternative formulation of a current source.) First, the logarithmic transformation of current exposes the small current region; the left panel in Figure \ref{hxfi} recapitulates \cite{duering_current_1992,duering_current_1990,straley_current_1989} the small currents.  Second, for relatively large currents (i.e., to the right of the peak of the $h(x)$ distribution), Figure \ref{hxfi} (right) is clearly suggestive of exponential current distributions. This general shape of the current distribution persists as bond density $p$ increases above threshold for a ``distance" $p-p_c$ to be clarified below relative to persistence of the small current scaling. The implications are two-fold: an exponential large current tail at percolation that persists for $p-p_c$ at finite non-zero values; a small current tail that disintegrates relatively rapidly above the critical bond density. We are cautious to rule out finite size effects since the $p=0.25$ correlation length \cite{stauffer_introduction_1994} is larger than the system size $L=100$. Specifically, near threshold, the correlation length $\xi$ scales as $\xi\propto (p-p_c)^{-0.9}$ \cite{stauffer_introduction_1994}, a feature that we will incorporate in our study
below.

While the small-current behavior is well understood and the maximum current has been studied extensively, little is known or has been reported about the large current tail of the distribution --- even though the large currents dominate bulk properties (which we explicitly show below). This is not so surprising in retrospect. The theoretical allure of the physics community has focused primarily on universality of power law scaling at percolation threshold, which is revealed by the small current tail as one approaches criticality from above. On the other hand, concerns of network failure draw attention to the largest current in the network. The bridge between these scaling behaviors (i.e., the large current tail), and the relative robustness of both tails of the distribution above percolation, do not appear to be addressed in the literature (with the caveat noted earlier of a cut-off function in analysis of the cluster size distribution). We turn attention to these issues next.

\subsection{Finite Size Scaling Analysis}
\begin{figure*}
  \begin{minipage}{0.5\textwidth}
  \includegraphics[width=\textwidth]{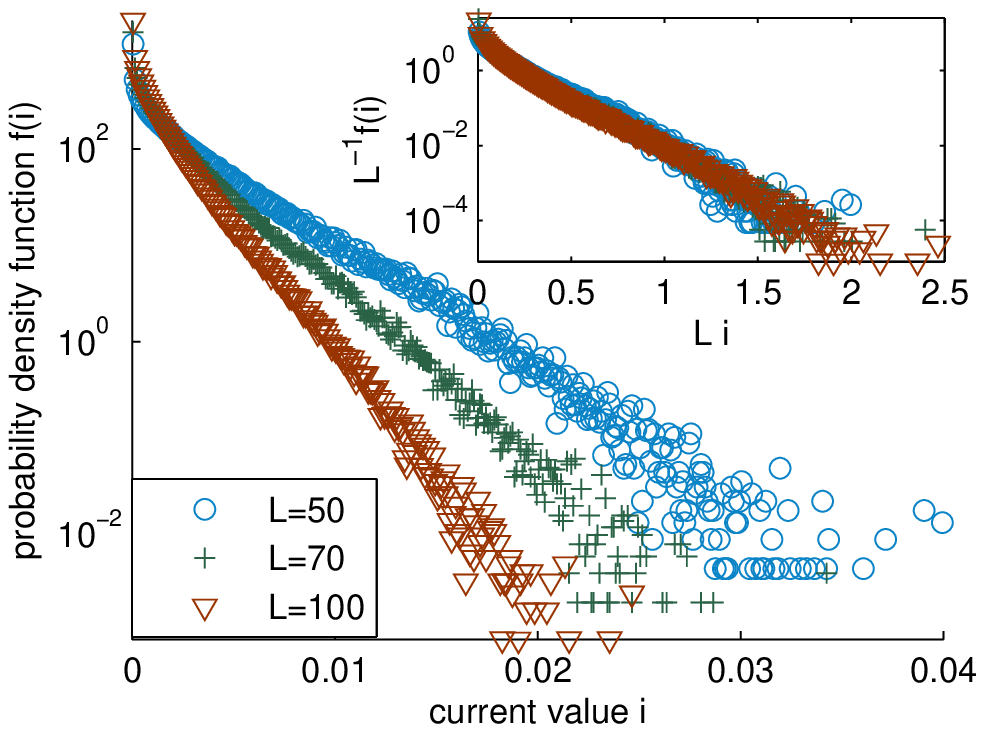}
  \end{minipage}%
  \begin{minipage}{0.5\textwidth}
  \includegraphics[width=\textwidth]{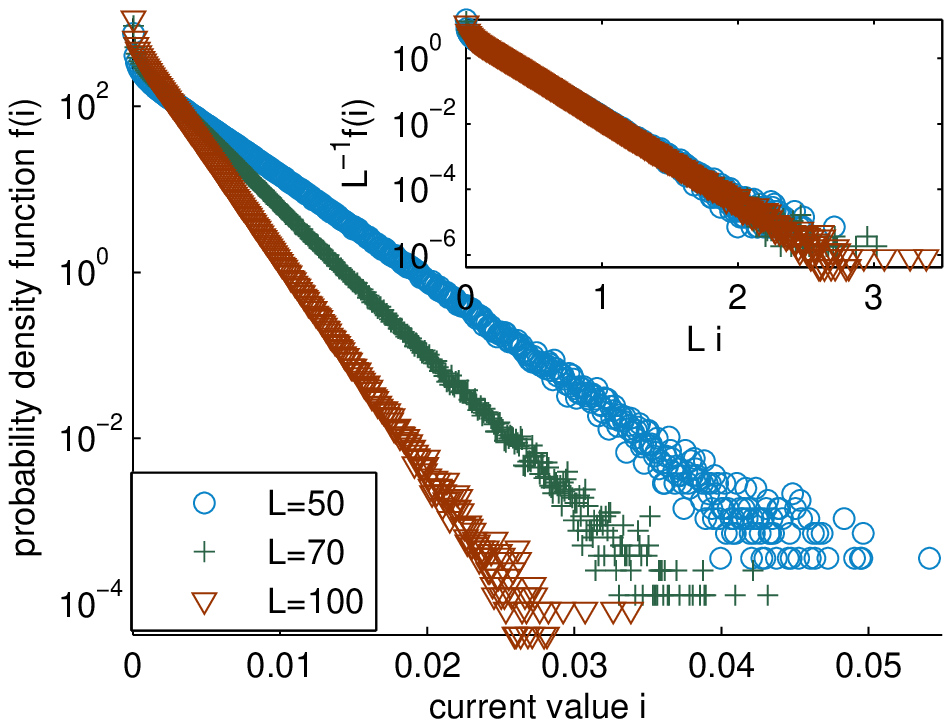}
  \end{minipage}
  \caption{(Color online) Empirical probability density function of the current in an $L\times L\times L$ cubic lattice at bond density $0.25$ (left) and at bond density 0.29 (right) for various system sizes $L$. The percolation threshold is at bond density $0.2488$. A constant unit voltage is imposed across the system. The plot is derived from the histograms of all the currents over $1000$ realizations. The inset shows the same distributions rescaled by the system $L$ with critical exponents $u=1$ and $v=1$.
  }\label{currentdistributions}
\end{figure*}

We now focus on the large current tail, revealing a robust exponential distribution above and close to percolation threshold, yet persistent farther from threshold than the small current power law scaling. The empirical densities of the current for a unit voltage source near percolation threshold ($p=0.25$) and above threshold ($p=0.29$) for different system sizes are shown in Figure \ref{currentdistributions}. Despite slightly larger noise at bond density $p=0.25$, the straight lines at both bond densities point to exponential tails of the current distributions, and the rate of the exponential decay increases with the system size $L$. In order to meaningfully capture an externally-imposed voltage drop in the thermodynamic limit ($L\rightarrow\infty$), and to better understand the effect of system size on the current distribution, we carry out a finite-size scaling analysis on the distributions. Let $f_L(i)$ be the probability density function (PDF) of the current at system size $L$ for a unit voltage source; then by properly rescaling $f_L(i)$ with $L$, we aim to eliminate the effect of the system size:
\begin{equation}\label{pdfil}
    L^{-u}f_L(L^{-v}i) =f^{\infty}(i),
\end{equation}
where $f^{\infty}(i)$ is a function independent of $L$, and $u$ and $v$ are critical
exponents to be determined. By tuning $u$ and $v$, we confirm that the densities for different system sizes collapse onto a single curve with $u=1$ and $v=1$ (see the insets of Figure \ref{currentdistributions}). This is the expected result for a material with bulk conductance: the total resistance of the cube per unit cross-sectional area increases $\sim L$. This results in $v=1$, while $u=1$ yields the correct normalization factor so that the rescaled PDF integrates to $1$. In other words, $f^{\infty}(i)$ is the limiting current distribution for system size $L$ and external voltage source $V^*=L$ (i.e., a unit uniform electric field, up to edge effects). Therefore in a finite system the current density for a unit voltage source scales as:
\begin{equation}\label{pdfil2}
    f_L(i) = L\, f^{\infty}(L i).
\end{equation}%

\begin{figure}
\centering
    \includegraphics[width=0.5\textwidth]{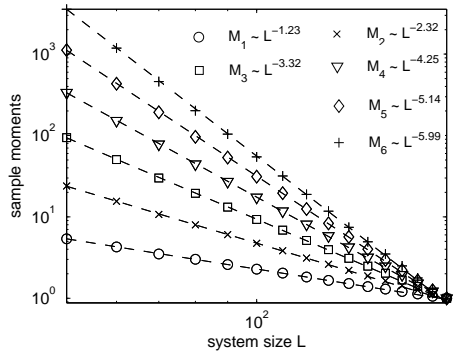}\\
    \caption{Scaling of the first $6$ sample moments with respect to the system size $L$. The bond density is fixed at $p=0.25$ and the system size $L$ varies from $50$ to $200$. The fitted equations of the moments $M_{k}$ are shown in the figure (cf. Equation (\ref{mk})). Curves are normalized by their values at $L=200$.}\label{momentsp025}
\end{figure}

It might seem natural that the current distribution will not change with system size $L$ if the electric field in the system is kept constant as $L$ increases. However, the simple scaling form in equation (\ref{pdfil2}) is not trivial. It implies that the multifractal property of the current distribution \cite{de_arcangelis_multiscaling_1986} comes from small currents, since the large current tail has a simple scaling form with respect to the system size. Specifically, the $k^{th}$ moment $M_k$ of the large currents described by this finite-size scaling is a simple scaling function of $L$:
\begin{equation}\label{mk}
    M_k = \int_0^{\infty}i^kf_L(i)\,di = \int_0^{\infty}i^kL\, f^{\infty}(L i)\,di \propto L^{-k}.
\end{equation}
To confirm this simple scaling form of the moments, we compute the first several sample moments at bond density $0.25$ for varying system sizes and plot them in Figure \ref{momentsp025}. The sample moments are calculated as $\hat{M}_k=\frac{1}{N}\sum_b i_b^k$, where $N$ is the number of bonds with nonzero current and the sum of bond currents $i_b$ is taken over all current-carrying bonds, $b$. The scaling forms of the moments are not exactly the same as Equation (\ref{mk}) due to the multifractal property of small currents and numerical error; however for large moments the exponential tail of the current distribution becomes dominant and thus the scaling relationship approaches Equation (\ref{mk}).

\subsection{Robustness of the Exponential Current Tail above Criticality}
\begin{figure}
\centering
  \includegraphics[width=0.5\textwidth]{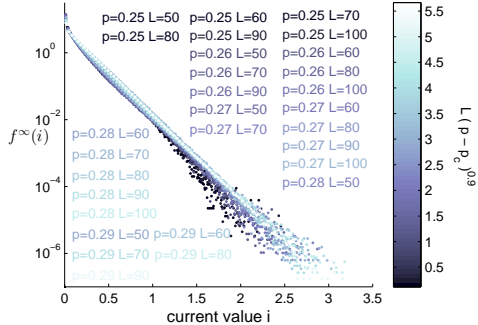}\\
  \caption{(Color Online) Limiting current distributions $f^{\infty}(i)=L^{-1}f_L(L^{-1}i)$ at different system sizes and bond densities. Colors represent the ratio $L/(p-p_c)^{-0.9}$ where $p_c\doteq0.2488$ is the percolation threshold for an infinite system and thus the color serves as an indicator of the system size relative to the correlation length. The overlap of current distributions at different parameters demonstrates a robust exponential distribution for large currents.}
  \label{allinone}
\end{figure}

Given the simple scaling form of the large current distribution with respect to system size, we now turn to the robustness of the exponential tail of the current distribution above criticality and its scaling with bond density $p$. We superimpose the limiting current distributions $f^{\infty}(i)=L^{-1}f_L(L^{-1}i)$ for a wide range of system sizes $L$ and bond densities $p$ in Figure \ref{allinone}, with colors representing the ratio $L/(p-p_c)^{-0.9}$ as an indicator of the extent to which our result is affected by the finite size effect. Large values of this ratio indicate that our system size is larger than the correlation length and hence have limited finite size effect. Despite the noise at low system-size-to-correlation-length ratios, {\em Figure \ref{allinone} demonstrates apparent convergence to a robust class of exponential distributions for the large current tails both near and above threshold. The rate of the exponential tail does not depend on the bond density and the simple scaling form of the tail of the distribution with respect to the system size remains the same}. We note that this robust feature of current distributions only holds sufficiently close to threshold. For instance, at saturation (p=1) the current distribution approaches a delta function which is the current distribution at $p=1$ where every bond on the straight-line paths perpendicular to the two plates carries the same current while all other bonds carry zero current.

To quantitatively confirm the independence of the exponential tail on the bond density $p$ given a unit voltage source, we now consider a unit current source flowing between the two plates. Current distributions for a unit voltage source differ from those for a unit current source by a factor of the bulk conductance $C$, because of the linearity of the system. Formally, denoting the density of the current distribution for the unit current source by $g_L(i)$, the density $f_L(i)$ of the current distribution for a unit voltage source can be written as:
$f_L(i)=g_L(i/C)/C$. Then assuming
$g_L(i)\sim e^{-\lambda(p) i}$ for large currents yields
\begin{equation}\label{fl}
    f_L(i)\sim\exp\{-\frac{\lambda(p)}{C(p)}i\}.
\end{equation}
Therefore, the rate of the exponential tail of $f_L(i)$ being independent of the bond density (over an observed range) indicates a linear relationship between the rate $\lambda(p)$ of decay of the exponential tail of $g_L(i)$ and the bulk conductance $C(p)$, and vice versa.

\begin{figure}
\centering
  \includegraphics[width=0.5\textwidth]{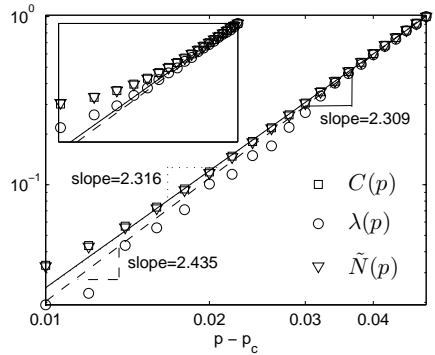}\\
  \caption{Scaling behaviors against bond density $p-p_c^\mathrm{eff}$ for the bulk conductance $C(p)$, the rate $\lambda(p)$ of the exponential tail of the current distribution, and the number $\tilde{N}(p)$ of bonds with current larger than $10^{-3}$. The system size is fixed at $L=100$ and the bond density $p$ varies from $0.25$ to $0.29$. Curves are normalized by their values at $p=0.29$. In the parent figure $p_c^\mathrm{eff}=0.24$ as defined in \cite{stauffer_universality_1999} and $p_c^\mathrm{eff}=0.2488$ in the inset. The inset demonstrates the sensitivity of the scaling exponent on the value of $p_c$ used (with fitted exponents $1.919\pm.012$, $1.969\pm.077$ and $1.917\pm.013$ respectively), but further supports the similar behavior of $\lambda(p)$ and $C(p)$.}
  \label{scaling2}
\end{figure}

It would seem surprising for these two properties to scale linearly with one another. To quantify this relationship, exponential distributions are fitted to the current distributions for a unit current source at various bond densities and their rates $\lambda(p)$ are plotted against the bond densities $p$ in Figure \ref{scaling2}, along with the scaling behavior of the bulk conductance $C(p)$. It is well known in percolation theory that the bulk conductance has a power-law scaling with respect to $p$  \cite{clerc_electrical_1990}; therefore $\lambda(p)$ also scales with $p$ as a power law. These power law scalings are confirmed in Figure \ref{scaling2}. Using the percolation threshold $p_c=0.2488$ of an infinite 3D cubic lattice  \cite{clerc_electrical_1990}, we find the conductivity exponent to be $1.919\pm0.012$ (see the inset of Figure \ref{scaling2}), agreeing with the literature values \cite{clerc_electrical_1990}. Moreover, since our system is finite we also consider the scaling behaviors with respect to $p-p_c^\mathrm{eff}(L)$ where $p_c^\mathrm{eff}(L)$ is defined as the effective percolation threshold for a finite system in Stauffer {\em et al.} \cite{stauffer_universality_1999}. Figure \ref{scaling2} demonstrates that the quantitative details of these power-law scalings are sensitive to the choice of $p_c^\mathrm{eff}(L)$, yet the point of emphasis here is that for either choice taken, there is a persistent linear relationship between the rate of exponential decay $\lambda(p)$ and the bulk conductance $C(p)$. These results indicate that $\lambda(p)$ and $C(p)$ have similar scaling behaviors with respect to bond density near and above the percolation threshold.

Both the underlying exponential distribution and the power-law scalings break down far above threshold, e.g., for $p\geq0.35$ as shown in the Appendix.

\subsection{Scaling Behaviors of the Largest Current}
Given the explicit exponential form of the tail of the current distribution, we calculate the statistics of the largest current in the system and show that they scale logarithmically with respect to system size $L$, agreeing with the literature \cite{chan_large_1989,kahng_logarithmic_1987,li_size_1987,machta_largest_1987}. Let $M_n$ be the largest current in the network with $n$ bonds and $F_{M_n}(i)$ be the cumulative distribution of $M_n$. Then assuming weak dependence between bonds and an exponential tail of the current distribution $f_L(i)\sim \exp(-i)$, we can write $F_M(i)$ asymptotically as
\begin{equation}\label{FM}
    F_{M_n}(i)\sim (1-e^{-i})^n
\end{equation}
for large current $i$. Therefore the mean of $M_n$ is
\begin{eqnarray*}
        \langle M_n\rangle &\approx& \int_0^{\infty} i\cdot n(1-e^{-i})^{n-1}e^{-i}di\\
                  &=& \sum_{\mu=1}^n 1/\mu \sim \ln n + \gamma.
\end{eqnarray*}
Substituting in $n=pL^3$, we recover the logarithmic scaling of the mean largest current $\langle M_n\rangle \sim \ln L$. Similarly, solving $F_{M_n}''(i)=0$ we find that the mode of $M_n$ is also $\ln n$, and the characteristic largest value \cite{gumbel_statistics_2004}, which is the $(n-1)^{\mathrm{th}}$ n-quantile of the current distribution, is calculated to be $\ln n$ as well. We note that the scaling of the largest value of a distribution does not generally imply the shape of the distribution, since it is not hard to construct different distributions with the same largest current scaling behavior. However, the current distribution in this model results from a normal physical system and hence it is expected to have a regular tail. Li {\em et al.} \cite{li_size_1987} showed that the logarithmic scaling of characteristic largest value would imply an  exponential tail of the current distribution. In the present work, we have not only identified an exponential tail but, more strongly, have demonstrated that this exponential behavior dominates the current distribution and the resulting macroscopic properties. Moreover, we have identified the dependence of that exponential on bond densities in a range above the percolation threshold.

\subsection{Large Current Tail and Scaling of the Bulk Conductance}
Bulk conductance of the random resistor network has been studied extensively in percolation theory \cite{clerc_electrical_1990, grimmett_percolation_1999, sahimi_applications_1994, stauffer_introduction_1994}. We have confirmed the power law scaling of the conductance and its exponent above (Figure \ref{scaling2}). We now connect the above scaling results for the current distribution to the reported power-law scaling of the macroscopic conductance. (In particular, we will show that the power law scaling in the small current tail of the current distribution at percolation threshold is not responsible for the power law scaling in bulk conductance.) Recall that the second moment of the current distribution is related to the conductance of the network. We now show that the scaling behavior of bulk conductance is inherited from the large current tail of the distribution. Specifically, conservation of energy in the system gives,
\begin{equation}\label{conservationlaw}
    \frac{V^2}{R}=\sum_b i_b^2 r_b,
\end{equation}
where $V$ and $R$ are the external voltage and the bulk resistance of the system respectively, $i_b$ is the value of the current on a bond, $r_b=1$ is the resistance of a bond, and the sum is taken over all bonds with nonzero current. For a unit external voltage source, Equation (\ref{conservationlaw}) can be rewritten as $C=R^{-1}=\sum_b i_b^2$, where $C$ is the bulk conductance of the network. Dividing by the number of current-carrying bonds $N$, we recover the conductance from the continuous current distribution:
\begin{equation}\label{continuum}
    \frac{C}{N}=\frac{\sum i_b^2}{N}=\int_0^{\infty} i^2f_L(i)\,di.
\end{equation}
Equation (\ref{continuum}) connects the scaling behavior of the bulk conductance to that of the current distribution. Since the second moment of the current distribution $f_L(i)$ is dominated by the exponential tail of $f_L(i)$ which is shown to be independent of the bond density $p$, the number $\tilde{N}(p)$ of bonds carrying large currents (whose magnitudes are assumed to be larger than $10^{-3}$) and the bulk conductance $C(p)$ have the same scaling form with respect to the bond density $p$, as demonstrated in Figure \ref{scaling2}. {\em The balance of Equation (\ref{continuum}) conditioned on large currents reveals an intrinsic consistency between the power-law scaling of the conductance and the exponential large current tail of the current distribution.}

\section{Conclusion}
We have identified a robust exponential large current tail of the global current distribution in percolating, 3D random resistor networks for the boundary value problem of conductance between two infinite parallel plates. This feature persists above percolation threshold at bond densities for which the celebrated small current power law scaling has already disintegrated. In the supercritical regime above percolation threshold, it is precisely this range of currents that is most relevant for describing and diagnosing the macroscopic electrical response for materials applications. Our numerical simulations leverage a network graph representation, whereby a breadth-first search preconditioner removes a large fraction ($\sim90\%$) of the {\it a priori} zero current carrying bonds. This approach both reduces the linear system to allow larger system sizes at fixed computational cost, and gives significantly better accuracy for capturing scaling behavior of the small current tail.

We surmise there is a geometric scaling behavior in the percolating bond component that underlies this exponential large current tail, yet this remains an open question. This structure-property relationship is ripe for understanding a variety of material performance properties. One possibility is the dynamics of failure due to random small perturbations in the network.  It is interesting to probe the extent to which the electrical property analysis is a proxy for characterization of the geometry of the percolating bond network. The percolating backbone is where the large current is supported, and one can further analyze the geometry of the bonds supporting successive values in the large current tail.  This geometry, once understood,  presumably can be used to explore more general structure-property relationships.  For example, does the geometry of the percolating backbone imply mechanical properties? It is well-known in the engineering literature that ``mechanical percolation" occurs at much higher volume fractions than ``electrical percolation".  We currently are pursuing the possible connection between geometry of the percolating component(s) and other transport properties of a random resistor network, such as mechanical properties and stability of bulk conductances to random perturbations among the bonds. Finally, the existence of the large current tail in this classical random resistor network model appears not to have been addressed previously in the literature, even though it is a dominant feature for theoretical and practical considerations. The exponential cut-off that has been explored in the cluster size distribution \cite{rubinstein_polymer_2003} may very well be related since it corresponds to the geometry of the percolating component, as suggested by our colleague Michael Rubinstein.

Partial research support has been provided by NSF DMR-1122483, DMS-1100281 and DMS-0645369, AFOSR FA9550-12-1-0178 and ARO 12-60317-MS.

\Appendix
\section{Scaling behaviors far from threshold}
The bond percolation threshold $p_c$ for an infinite 3D cubic lattice is $p_c\doteq 0.2488$ \cite{lorenz_precise_1998}. Near and above threshold we have identified power-law scaling behaviors of various macroscopic electrical properties such as the bulk conductance and the number of bonds with nonzero current, and a robust and universal exponential distribution describing large currents. We note that both the power-law scalings and the underlying exponential distributions become invalid further from the threshold as the system saturates.

\begin{figure}
\centering
  \includegraphics[width=0.5\textwidth]{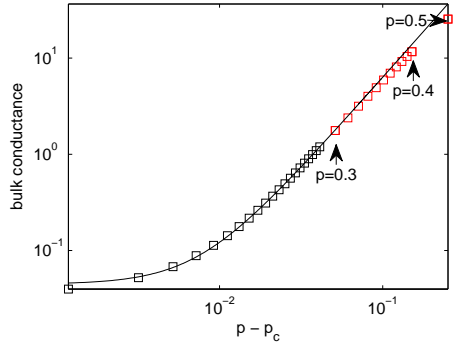}\\
  \caption{Scaling of the bulk conductance with respect to the bond density $p$ in an $100\times 100\times 100$ cubic lattice.}\label{conductance_largep}
\end{figure}

\begin{figure}
\centering
  \includegraphics[width=0.5\textwidth]{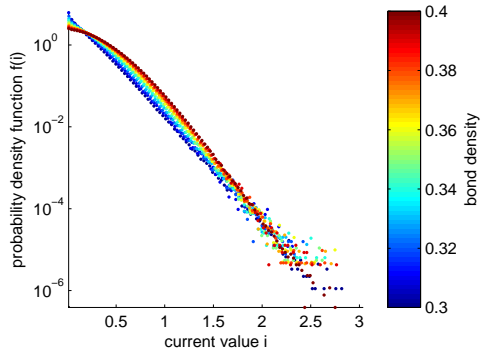}\\
  \caption{(Color online) Limiting current distributions at bond density ranging from $p=.3-.4$.}\label{fi0_3_0_4}
\end{figure}

Figure \ref{conductance_largep} plots the bulk conductance for bond densities up to $.5$. The bulk conductance gradually deviates from the power-law scaling near threshold after $p=.3$. Figure \ref{fi0_3_0_4} shows the current distributions at large bond densities with colors representing the bond density $p$, where $p$ ranges from $.3$ to $.4$. The current distribution at $p=.3$ still agrees with the universal exponential distribution while as $p$ increases it approaches a delta function which is the current distribution at $p=1$ where every bond on the paths perpendicular to the two plates carries the same current while all other bonds carry zero current.


\bibliographystyle{plain}
\bibliography{mybib}

\begin{thebibliography}{10}

\bibitem{balberg_excluded_1984}
I.~Balberg, C.~H. Anderson, S.~Alexander, and N.~Wagner.
\newblock Excluded volume and its relation to the onset of percolation.
\newblock {\em Physical Review B}, 30(7):3933--3943, October 1984.

\bibitem{batrouni_fracture_1998}
G.~George Batrouni and Alex Hansen.
\newblock Fracture in three-dimensional fuse networks.
\newblock {\em Physical Review Letters}, 80(2):325--328, January 1998.

\bibitem{bauhofer_review_2009}
W.~Bauhofer and J.~Z. Kovacs.
\newblock A review and analysis of electrical percolation in carbon nanotube
  polymer composites.
\newblock {\em Composites Science and Technology}, 69(10):1486--1498, August
  2009.

\bibitem{bergman_physical_1992}
D.~J. Bergman and D.~Stroud.
\newblock Physical properties of macroscopically inhomogeneous media.
\newblock In {\em Solid State Physics}, volume~46, pages 147--269. 1992.

\bibitem{berhan_modeling_2007}
L.~Berhan and A.~M. Sastry.
\newblock Modeling percolation in high-aspect-ratio fiber systems. i. soft-core
  versus hard-core models.
\newblock {\em Physical review. E, Statistical, nonlinear, and soft matter
  physics}, 75(4 Pt 1):041120, April 2007.
\newblock {PMID:} 17500878.

\bibitem{bug_continuum_1985}
A.~L.~R. Bug, S.~A. Safran, and I.~Webman.
\newblock Continuum percolation of rods.
\newblock {\em Physical Review Letters}, 54(13):1412--1415, April 1985.

\bibitem{byrne_recent_2010}
M.~T. Byrne and Y.~K. Gun'ko.
\newblock Recent advances in research on carbon {Nanotube-Polymer} composites.
\newblock {\em Advanced Materials}, 22(15):1672--1688, 2010.

\bibitem{chan_large_1989}
S.-{K}. Chan, J.~Machta, and R.~A. Guyer.
\newblock Large currents in random resistor networks.
\newblock {\em Physical Review B}, 39(13):9236--9239, May 1989.

\bibitem{clerc_electrical_1990}
{J.P.} Clerc, G.~Giraud, {J.M.} Laugier, and {J.M.} Luck.
\newblock The electrical conductivity of binary disordered systems, percolation
  clusters, fractals and related models.
\newblock {\em Advances in Physics}, 39(3):191--309, 1990.

\bibitem{de_arcangelis_anomalous_1985}
L.~de~Arcangelis, S.~Redner, and A.~Coniglio.
\newblock Anomalous voltage distribution of random resistor networks and a new
  model for the backbone at the percolation threshold.
\newblock {\em Physical Review B}, 31(7):4725, 1985.

\bibitem{de_arcangelis_multiscaling_1986}
L.~de~Arcangelis, S.~Redner, and A.~Coniglio.
\newblock Multiscaling approach in random resistor and random superconducting
  networks.
\newblock {\em Physical Review B}, 34(7):4656, October 1986.

\bibitem{de_arcangelis_random_1985}
L.~de~Arcangelis, S.~Redner, and {H.J.} Herrmann.
\newblock A random fuse model for breaking processes.
\newblock {\em Journal de Physique Lettres}, 46(13):585--590, 1985.

\bibitem{deutscher_percolation_1983}
G.~Deutscher, R.~Zallen, and J.~Adler.
\newblock {\em Percolation structures and processes}.
\newblock A. Hilger, 1983.

\bibitem{duering_current_1990}
E.~Duering and D.~J. Bergman.
\newblock Current distribution on a three-dimensional, bond-diluted,
  random-resistor network at the percolation threshold.
\newblock {\em Journal of Statistical Physics}, 60(3-4):363--381, August 1990.

\bibitem{duering_current_1992}
E.~Duering, R.~Blumenfeld, D.~J. Bergman, A.~Aharony, and M.~Murat.
\newblock Current distributions in a two-dimensional random-resistor network.
\newblock {\em Journal of Statistical Physics}, 67(1-2):113--121, April 1992.

\bibitem{duxbury_breakdown_1987}
P.~M. Duxbury, P.~L. Leath, and P.~D. Beale.
\newblock Breakdown properties of quenched random systems: The random-fuse
  network.
\newblock {\em Physical Review B}, 36(1):367--380, July 1987.

\bibitem{foygel_theoretical_2005}
M.~Foygel, R.~D. Morris, D.~Anez, S.~French, and V.~L. Sobolev.
\newblock Theoretical and computational studies of carbon nanotube composites
  and suspensions: Electrical and thermal conductivity.
\newblock {\em Physical Review B}, 71(10):104201, March 2005.

\bibitem{grimmett_percolation_1999}
G.~R. Grimmett.
\newblock {\em Percolation}.
\newblock Springer, 2nd edition, June 1999.

\bibitem{gumbel_statistics_2004}
E.~J. Gumbel.
\newblock {\em Statistics of Extremes}.
\newblock Dover Publications, July 2004.

\bibitem{kahng_electrical_1988}
B.~Kahng, G.~Batrouni, S.~Redner, L.~de~Arcangelis, and H.~Herrmann.
\newblock Electrical breakdown in a fuse network with random, continuously
  distributed breaking strengths.
\newblock {\em Physical Review B}, 37(13):7625--7637, May 1988.

\bibitem{kahng_logarithmic_1987}
B.~Kahng, {G.G.} Batrouni, and S.~Redner.
\newblock Logarithmic voltage anomalies in random resistor networks.
\newblock {\em Journal of Physics A: Mathematical and General}, 20:L827, 1987.

\bibitem{kesten_percolation_1982}
H.~Kesten.
\newblock {\em Percolation Theory for Mathematics}.
\newblock Birkhäuser Boston, 1 edition, November 1982.

\bibitem{kirkpatrick_percolation_1973}
S.~Kirkpatrick.
\newblock Percolation and conduction.
\newblock {\em Reviews of Modern Physics}, 45(4):574--588, October 1973.

\bibitem{knuth_art_1997}
D.~E. Knuth.
\newblock {\em Art of Computer Programming, Volume 1: Fundamental Algorithms}.
\newblock Addison-Wesley Professional, 3 edition, July 1997.

\bibitem{li_size_1987}
Y.~S. Li and P.~M. Duxbury.
\newblock Size and location of the largest current in a random resistor
  network.
\newblock {\em Physical Review B}, 36(10):5411, October 1987.

\bibitem{lin_multifractal_1991}
B.~Lin, Z.-Z. Zhang, and B.~Hu.
\newblock Multifractal characterization of random resistor and random
  superconductor networks.
\newblock {\em Physical Review A}, 44(2):960, July 1991.

\bibitem{lorenz_precise_1998}
C.~D. Lorenz and R.~M. Ziff.
\newblock Precise determination of the bond percolation thresholds and
  finite-size scaling corrections for the sc, fcc, and bcc lattices.
\newblock {\em Physical Review E}, 57(1):230--236, January 1998.

\bibitem{machta_largest_1987}
J.~Machta and R.~A. Guyer.
\newblock Largest current in a random resistor network.
\newblock {\em Physical Review B}, 36(4):2142, 1987.

\bibitem{nan_physical_2010}
C.-W. Nan, Y.~Shen, and J.~Ma.
\newblock Physical properties of composites near percolation.
\newblock {\em Annual Review of Materials Research}, 40(1):131--151, June 2010.

\bibitem{neda_reconsideration_1999}
Z.~N\'{e}da, R.~Florian, and Y.~Brechet.
\newblock Reconsideration of continuum percolation of isotropically oriented
  sticks in three dimensions.
\newblock {\em Physical Review E}, 59(3):3717--3719, March 1999.

\bibitem{rammal_flicker_1985}
R.~Rammal, C.~Tannous, P.~Breton, and A.~M.~S. Tremblay.
\newblock Flicker (1/f) noise in percolation networks: A new hierarchy of
  exponents.
\newblock {\em Physical Review Letters}, 54(15):1718--1721, April 1985.

\bibitem{rubinstein_polymer_2003}
M.~Rubinstein and R.~H. Colby.
\newblock {\em Polymer physics}.
\newblock Oxford University Press, Oxford; New York, 2003.

\bibitem{sahimi_applications_1994}
M.~Sahimi.
\newblock {\em Applications of {P}ercolation {T}heory}.
\newblock Taylor \& Francis, London; Bristol, {PA}, 1994.

\bibitem{shi_network_2013}
F.~Shi, S.~Wang, M.~G. Forest, P.~J. Mucha, and R.~Zhou.
\newblock Network-based assessments of percolation-induced current
  distributions in sheared rod macromolecular dispersions.
\newblock {\em SIAM-MMS In Review}, 2013.

\bibitem{stauffer_universality_1999}
D.~Stauffer, J.~Adler, and A.~Aharony.
\newblock Universality at the three-dimensional percolation threshold.
\newblock {\em Journal of Physics A: Mathematical and General}, 27(13):L475,
  1999.

\bibitem{stauffer_introduction_1994}
D.~Stauffer and A.~Aharony.
\newblock {\em Introduction to {P}ercolation {T}heory}.
\newblock {CRC} Press, 1994.

\bibitem{straley_current_1989}
J.~P. Straley.
\newblock Current distribution in random resistor networks.
\newblock {\em Physical Review B}, 39(7):4531, March 1989.

\bibitem{strang_introduction_1986}
G.~Strang.
\newblock {\em Introduction to Applied Mathematics}.
\newblock Wellesley-Cambridge Press, January 1986.

\bibitem{white_electrical_2010}
S.~I. White, R.~M. Mutiso, P.~M. Vora, D.~Jahnke, S.~Hsu, J.~M. Kikkawa, Ju~Li,
  J.~E. Fischer, and K.~I. Winey.
\newblock Electrical percolation behavior in silver nanowire-polystyrene
  composites: Simulation and experiment.
\newblock {\em Advanced Functional Materials}, 20(16):2709--2716, 2010.

\end{thebibliography}

\end{document}